\documentclass[prl,twocolumn,preprintnumbers]{revtex4}
\usepackage{mathrsfs,graphicx,amssymb,amsbsy}

\def\scr{\mathscr}
   
  \def\SE{{\scr E}}

\def\Bid{{\mathchoice {\rm {1\mskip-4.5mu l}} {\rm
{1\mskip-4.5mu l}} {\rm {1\mskip-3.8mu l}} {\rm {1\mskip-4.3mu l}}}}

\def\avg#1{\langle#1\rangle}    \def\<{\langle}     \def\>{\rangle}

\def\al{\alpha}     \def\bt{\beta}      
\def\sig{\sigma}    \def\del{\delta}    \def\Del{\Delta}
\def\eps{\epsilon}    
\def\up{\uparrow}   \def\down{\downarrow}

\def\Bk{{\mathbf k}} \def\Bp{{\mathbf p}} \def\Bq{{\mathbf q}}
\def\BQ{{\mathbf Q}}     
\def\Bpi{\boldsymbol{\pi}}

  \def\B0{{\mathbf 0}}
\def\Br{{\mathbf r}}

\def\be{\begin{equation}}   \def\ee{\end{equation}}
\def\bea{\begin{eqnarray}}  \def\eea{\end{eqnarray}}
\def\nn{\nonumber}

\begin{document}
\title{Spin-Orbit Ordering, Momentum Space Coexistence, and Cuprate
Superconductivity}

\author{W. Vincent Liu}
\author{Frank Wilczek}
\affiliation{Center for Theoretical Physics, Department of Physics,
Massachusetts Institute of Technology,
Cambridge, Massachusetts 02139}

\date{\today}
\preprint{MIT-CTP-3436}

\begin{abstract}

Motivated by the energetic advantage of achieving coherent enhancement
of effective spin-dependent interactions through approximate nesting,
we propose specific forms of spin ordering, whose form varies over the
Fermi surface, for the cuprate superconductors.  Competing
``spin-orbit" orderings involving order parameters in spatial
$d_{x^2-y^2}$ and $d_{xy}$ waves at commensurate and incommensurate
wavevectors, phase separated in momentum space, support behavior
suggestive of observed phenomena.  The $d_{xy}$ spin-orbit fluctuation
induces an effective interaction that favors $d_{x^2-y^2}$-wave
pairing, as required for the observed superconductivity.  Anisotropic
spin susceptibility is a crucial prediction of our mechanism.

\end{abstract}

\maketitle



High $T_c$ superconductivity in cuprates has been an outstanding
problem in condensed matter physics since it was first discovered in
1986. 
In the meantime, though no consensus
theoretical understanding has developed, several striking features of
the phenomenology have emerged clearly.  These include the strongly
2-dimensional character of the essentially new physics; its proximity
to antiferromagnetic spin ordering; the anomalous character of the
normal state, especially in the underdoped region, suggestive of
emergent energy gaps; the acute sensitivity of the low-temperature
state to doping and impurities; the d-wave character of the
superconductivity; and the apparent uniqueness or near-uniqueness of
cuprate layers in supporting this overall phenomenological profile.  On the
face of it, several of these features suggest that the origin of these
phenomena will involve forms of ordering that involve spin and depend
sensitively on the precise form of the lattice and the Fermi surface.

There is a simple heuristic that appears to be broadly consistent with
these indications.  As is familiar from the BCS theory of
superconductivity, the effect of weak attractive interactions can be
amplified, and can lead to drastic qualitative effects, if there are
many low-energy pairs sharing the same quantum numbers.  Correlations
among these pairs can then be arranged so that their interactions
contribute coherently to lowering the energy.  The BCS mechanism of
superconductivity involves particle-particle and hole-hole
correlations.  In this context the existence of low-energy pairs with
total momentum zero, arising from time-reversed states with momenta
$(\Bk, -\Bk)$ both near the Fermi surface, is generic.  By contrast
spin-density-wave 
ordering, at the level of electron creation and destruction operators,
involves particle-hole correlations.  In this context one finds
that many low-energy pairs sharing a common (lattice) momentum only for
specially shaped (``nested'') Fermi surfaces.  Since the shape of the
Fermi surface changes with doping, one might anticipate that at best a
nesting condition would be approximately fulfilled at a specific
doping level.  Our point of departure is to consider that the Fermi
surface is not an end in itself, but a step toward constructing the
ground state.
If changing the
pattern of occupied levels --- effectively, engineering the Fermi
surface --- can encourage favorable coherence factors,
it might be favorable to make
nesting persist.  Realizations of this possibility and the
properties of the emergent states will, on the face of it, depend
sensitively on details of the interactions, lattice structure, and
doping level.  Two dimensional antiferromagnets on a square (or nearly
square) lattice near half filling, as in cuprate layers, provide an 
especially favorable
area for these ideas.

\begin{figure}[htbp]
\begin{center}
\includegraphics[width=\linewidth]{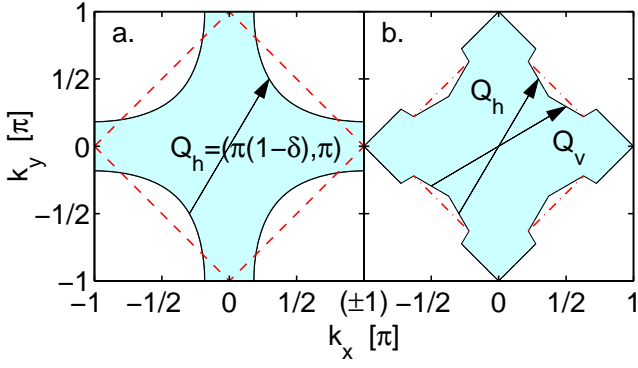}
\end{center}
\caption{Phase separation in momentum space. (a) Incommensurate
ordering at $\delta=2
\arcsin(|\mu|/2t)$~\cite{Schulz:90,Littlewood:93} is suggested by
approximate nesting for a Fermi surface with representative 
values of nearest and
next-nearest neighbor hopping, $t$ and $t'$,
respectively.  (b) A re-arrangement of occupied states allows 
incommensurate nesting to occur together with $\Bpi$ nesting for
finite doping.}
\label{fig:mom_phase_sep}
\end{figure}

\paragraph{Proposed Ordering} 
At half filling antiferromagnetic (AF) spin
ordering is observed.  Electrons in the real materials may well be
best described as strongly
coupled and the spins as localized, but we shall construct our states
heuristically by
extrapolation from weak or intermediate coupling, anticipating that
universal properties,
specifically symmetry breaking patterns, might be successfully
inferred. (Also, photoemission experiments~\cite{ARPES:rev:03} 
can be interpreted as 
revealing a Fermi surface, even at quite small doping.)  In that 
spirit AF
ordering can be regarded as follows.  On-site Coulomb repulsion
induces, in the crossed
channel, an attractive interaction between electrons and holes
of opposite spin
at momentum transfer
$\Bpi=(\pi,\pi)$ (modulo reciprocal lattice vectors).  This makes it
favorable to deform the
effective Fermi surface into a diamond shape, which for half filling
nests at $\Bpi$ (see
Fig.~\ref{fig:mom_phase_sep}a), and allow the electrons to form
spin-triplet particle-hole
pairs, according to $\avg{c^\dag_{\Bk+\Bpi}
\vec{\sig} c_{\Bk}} \neq 0$. Here, $\vec{\sig}=(\sig_1,\sig_2,\sig_3)$
are Pauli matrices, and $c_\Bk$ and
$c^\dag_\Bk$ are  electron annihilation and creation operators, with
two spin components united into a column:
$c^\dag_\Bk=(c^\dag_{\Bk\up}, c^\dag_{\Bk\down})$.

At finite doping no {\it ansatz\/} seems so uniquely compelling,
but the possibility illustrated in
Fig.~\ref{fig:mom_phase_sep}b is suggestive.  The free Fermi surface has been
deformed in two distinct ways, one operating near to and the other far
from the zone diagonals.
This geometry supports nesting for
orderings with wavevector $\Bpi$ in the off-diagonal region and with
wavevectors slightly off $\Bpi$,
namely, $\pm\BQ_h=((1\mp \delta)\pi,\pi)$ and $\pm\BQ_v=(\pi,(1\mp
\delta)\pi)$, along the diagonals. This proposal embodies a new phenomenon,
phase separation in momentum
space, that might find wider application.

Order parameters with near-uniform $s$-wave structure do not easily
accommodate such
phase separation.  It is more natural when zeroes of one condensate
correspond to maxima
of the other.  This leads us to propose
spin-orbit (SO) orders of the form:~\cite{remark:dSDW}
\bea
d_{x^2-y^2}: &&
\avg{c^\dag_{\Bk+\Bpi}\vec{\sig} c_\Bk} =
     {i} \vec{N}_\mathrm{A} \Gamma^A_\Bk\,,  \label{eq:dx2y2SDW}\\
     && \mbox{($\Bk\in$ off-diagonal)} \nn \\
d_{xy}: &&
\avg{c^\dag_{\Bk+\Bpi\pm {\pi\del }\hat{e}_j }\vec{\sig} c_\Bk} =
    \vec{N}_\mathrm{B} \Gamma^B_{\Bk\pm {\pi\del\over 2}\hat{e}_j } \,,
    \label{eq:dxySDW} \\
    && \mbox{($j=x,y$; $\Bk\in$ diagonal)} \nn
\eea
Here
$\vec{N}_\mathrm{A, B}$ are
constant, real vectors in spin space, and
$\Gamma^{A,B}_\Bk$ are orbital wave basis functions of lattice group
$D_{4h}$, defined by
$\Gamma^A_\Bk \equiv  \cos k_x -\cos k_y$ and
$ \Gamma^B_\Bk \equiv \sin k_x \sin k_y$. (The lattice constant is set
to a unit.) 
The $d_{x^2-y^2}$ order is purely imaginary, as required by
hemiticity.  Roughly speaking, it describes a state with microscopic
spin currents flowing
around each plaquette in real space.

The nature of these spin-orbit orderings may be more
transparent in real space:
\bea
\avg{c^\dag_\Br \vec{\sig} c_{\Br'}} &=&
    i\vec{N}_\mathrm{A}
     e^{i\Bpi\cdot \Br} \Gamma^A_{\Br-\Br'}
+ 2\vec{N}_\mathrm{B} e^{i\Bpi\cdot \Br} \nn \\
&& \times \textstyle
[\cos\left({x+x'\over 2} \pi \del\right) + (x\rightarrow y)]
\Gamma^B_{\Br-\Br'}  \label{eq:SOWr}
\eea
where
$
\Gamma^A_{\Br''} = {1\over 2} [(\del_{\Br'',\hat{x}} +\del_{\Br'',-\hat{x}}) -
     (\hat{x}\rightarrow \hat{y})  ]$ and
$\Gamma^B_{\Br''} = -{1\over 4} [\del_{\Br'',\hat{x}+\hat{y}} -
\del_{\Br'',\hat{x}-\hat{y}}-\del_{\Br'',-\hat{x}+\hat{y}}
+\del_{\Br'',-\hat{x}-\hat{y}}] \,.
$
Note that these vanish for $\Br= \Br'$.

At mean field level such spin-orbit orderings with nonzero $N_A$ and
$N_B$ are
favored by the nearest and next nearest neighbor Coulomb
repulsions, respectively.
Indeed
the nearest-neighbor  Coulomb interaction
$
H_V= + V\sum_{\avg{\Br\Br'}} n_\Br n_{\Br'}
$
can be transformed to
\be \textstyle
    -{V\over 2} \sum_{\avg{\Br\Br'}}\Big[ (c^\dag_{\Br} c_{\Br'})
       (c^\dag_{\Br'} c_{\Br})
        +
     (c^\dag_{\Br}\vec{\sig}c_{\Br'}) \cdot
      (c^\dag_{\Br'}\vec{\sig}c_{\Br})\Big] \,
\ee
plus an unimportant term proportional to the density operator.  In
this form the anticipated
electron-hole attraction is manifest.
(At this level $H_V$, with
$V>0$, also favors charge-orbit (CO) ordering, known as
orbital-antiferromagnetism, 
staggered flux phase~\cite{Affleck-Marston:88;Hsu+:91}, or
DDW~\cite{Chakravarty+:01}; a non-static version was proposed in
\cite{Wen-Lee:96}.
We shall not
discuss it further here.)
Similarly, the next-nearest-neighbor Coulomb
interaction favors both $d_{xy}$ SO and CO
orders.

\paragraph{Phase diagram}

\begin{figure}[htbp]
\begin{center}
\includegraphics[width=0.8\linewidth]{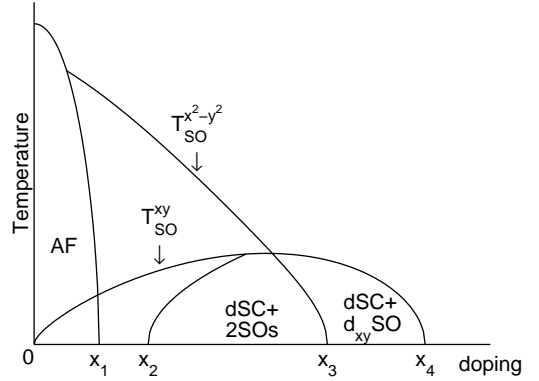}
\end{center}
\caption{Schematic phase diagram for cuprate superconductors.  The
    doping dependence of the transition temperatures
    $T^{x^2-y^2}_\mathrm{SO}$ and $T^{xy}_\mathrm{SO}$ results from
    the competition of the two SO orders in phase space (see
    Fig.~\ref{fig:mom_phase_sep}b). The doping value $x_2$ for the
    onset of dSC is estimated as the level of doping above which the
    Fermi surface near $(\pm\pi,0)$ and $(0,\pm\pi)$ starts to
    appear. To the right and from above, the dSC state is bounded by
    the $d_{xy}$-SO order, upon which it depends.}
\label{fig:phase_diag}
\end{figure}
We shall focus on the following
competing orders, that we believe play major roles:
AF, $d_{x^2-y^2}$-SO and $d_{xy}$-SO, and 
$d_{x^2-y^2}$-wave superconducting state
(dSC).  The two SO orders compete for particle-hole pairing 
states, plausibly with different outcomes in different domains, whose 
size depends on overall doping level.
This can be understood by reference to Fig.~\ref{fig:mom_phase_sep}b.  With
increasing density $x$ of doped holes, the total area of the four 
triangles along
the zone diagonals increases at first while the size of off-diagonal regions
shrinks. For sufficient large doping, e.g., $x_4$,
the triangles are no long
sustainable.  
This explains the trends of phase transition lines of
$T^{x^2-y^2}_\mathrm{SO}$  (for $d_{x^2-y^2}$-SO) and
     $T^{xy}_\mathrm{SO}$ (for $d_{xy}$-SO) as functions of doping.

For low doping up to $x_2$,  there are several competing
spin orders (including AF)  
that are connected by first-order phase transitions.  On
general grounds, one expects that phase separation (in real space)
may occur, plausibly in the form of stripes
\cite{Emery-Kivelson-Tranquada:99,Zaanen:00pre}).


\paragraph{Effective Theory}
\label{sec:EFT}
We now propose an effective Lagrangian for the unconventional normal 
state, based on the hypothesis of $d_{x^2-y^2}$-SO ordering:
\bea
L &=& \sum_\Br c^\dag_{\Br\sig} (i\partial_t+\mu) c_{\Br\sig} -H_0
   +{V_A}\sum_{\Br}\vec{\phi}_{\Br}\cdot \vec{\phi}_{\Br}
   \label{eq:H0V}
\eea
where $\vec{\phi}_\Br \equiv {1\over 2i} \sum_{\Br'} \Gamma^A_{\Br-\Br'}
[c^\dag_\Br\vec{\sig}  c_{\Br'} -\mbox{h.c.}]$ is a composite
operator, $V_A>0$ is assumed, and $H_0$ collects the hopping terms.
The interaction term
contains part of the original next-nearest neighbor Coulomb interaction.

A $d_{x^2-y^2}$-SO ordered state is characterized by
a non-vanishing order parameter
field $\vec{\phi}$, corresponding to the imaginary part of
$c^\dag_\Br\vec{\sig} c_{\Br'}$, pointing to (say) the
$3$-direction in spin space,
\be
\avg{\phi^{1,2}_{\Br}}=0\,, \quad
\avg{{\phi}^3_{\Br}} = e^{i\Bpi\cdot \Br} |\vec{N}_\mathrm{A}|\neq 0\,.
\label{eq:<phi>}
\ee
This correlation spontaneously breaks SU(2) spin symmetry down to a U(1)
corresponding to
spin rotation about the $3$-axis. Two Nambu-Goldstone
bosons appear as gapless collective spin excitations,
corresponding to transverse  fluctuations of the order
parameter field $\vec{\phi}_{\Br}$. We will call them
orbital magnons.

We can describe the low energy interactions of these collective modes using an
effective Lagrangian.  Following the standard technique
~\cite{Weinberg:bk96:ch19}, we isolate
the Nambu-Goldstone part (two transverse spin fluctuations) in the
electron field.  Expressing
the electron field as a local
SU(2) spin rotation acting on a new fermion $\psi$:
\bea
c_{\Br} = {\Bid +i\eps_{\al\bt}\sig_\al
\zeta_{\bt\Br}\over
    \sqrt{1+\vec{\zeta}_{\Br}^2}}
\, \psi_{\Br} \equiv U_\Br \psi_{\Br}
\,,  \quad \mbox{($\al,\bt=1,2$; no $3$)}.
   \label{eq:cb}
\eea
This defines a new fermion $\psi_{\sig\Br}$ that carries the full
charge and the spin $\sig_3$ quantum numbers
but does not carry transverse spin.
$\zeta^{1,2}$ parameterize the slowly-varying
orbital magnons. They are given by
the $3$-component $\vec{\phi}$ via: $U^\dag_\Br \sig^a U_\Br
=R_{\Br}^{ab}\sig^b$ and $\phi^a_{\Br}
= R^{a3}_{\Br} (\avg{\phi_3} +\varrho_\Br)$ with $a,b=1,2,3$, 
where $\varrho$ represents
the (gapped) longitudinal spin-orbit fluctuation.
With the above transformations, we now have a prescription of deriving
an effective Lagrangian for the orbital magnons
and new fermions: $L[\psi, \psi^\dag,\zeta^{\al},\varrho]$ from the Lagrangian
(\ref{eq:H0V}).

\begin{figure}[htbp]
\begin{center}
\includegraphics[width=0.5\linewidth]{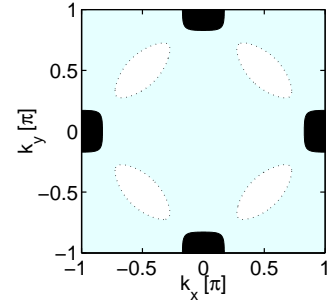}
\end{center}
\caption{The Fermi surface in $d_{x^2-y^2}$-SO state.  $E^-_\Bk=0$
    is shown as the dotted ellipses, inside which all states are
    empty.  $E^+_\Bk=0$ is shown as the boundaries of the four patches
    (in dark) which are completely occupied.  }
\label{fig:fermi_surf_dx2y2}
\end{figure}

The free part of the effective
theory for the $\psi$-fermion is
\be\begin{array}{rl}
L_\psi = \sum_\Bk \Big\{& \psi^\dag_\Bk [i\partial_t-\eps(\Bk) +\mu]
\psi_\Bk  \\
&   + i|\vec{\Del}_\Bk| [\psi^\dag_{\Bk+\Bpi} \sig^3 \psi_\Bk -
\mathrm{h.c.}]\Big\}
\end{array} \label{eq:Lpsi}
\ee
where $\eps_\Bk= -2t(\cos k_x +\cos k_y) - 4t'\cos k_x \cos k_y$ and
$\vec{\Del}_\Bk=\vec{\Del}_A (\cos k_x - \cos k_y)$ with
$\vec{\Del}_A = {1\over 2} V_A \vec{N}_\mathrm{A}$.
Due to unit cell doubling, the fermion energy spectrum is split into two
bands
\be\textstyle
E^{\pm}_\Bk = {1\over 2}(\eps_{\Bk}+\eps_{\Bk+\Bpi}) \pm
\sqrt{{1\over 4}(\eps_{\Bk}-\eps_{\Bk+\Bpi})^2 +|\vec{\Del}_\Bk|^2}\,,
\ee
each being spin up and down degenerate.
For weak $d_{x^2-y^2}$-SO order, the Fermi surface of the $E^-$ band 
is near the diagonals of the Brillouin zone while
the Fermi surface of the $E^+$ band is near the off-diagonal region 
(see Fig.~\ref{fig:fermi_surf_dx2y2}).
Note that the fermion field $\psi$, appropriate to describing the 
low-energy excitations, does not carry the full spin quantum numbers 
of the electron.  This is a form of partial spin-charge separation.

The effective theory for $d_{xy}$-SO can be constructed in the same
manner from a model Lagrangian similar to (\ref{eq:H0V}), with
$V_A\rightarrow V_B$ and $\vec{\phi}_\Br \rightarrow   
\vec{\phi}_\Br \equiv 2\sum_{\Br'} \Gamma^B_{\Br-\Br'}
[c^\dag_\Br\vec{\sig}  c_{\Br'} +\mbox{h.c.}]$ 
($d_{xy}$-SO order parameter). The ordering wavevector is changed from
$\Bpi$ to $\BQ_{h,v}$.

\paragraph{Origin of $d$-wave Superconductivity}

\begin{figure}[htbp]
\begin{tabular}{cc}
\raisebox{2em}{\includegraphics[width=.4\linewidth,
height=0.4\linewidth]{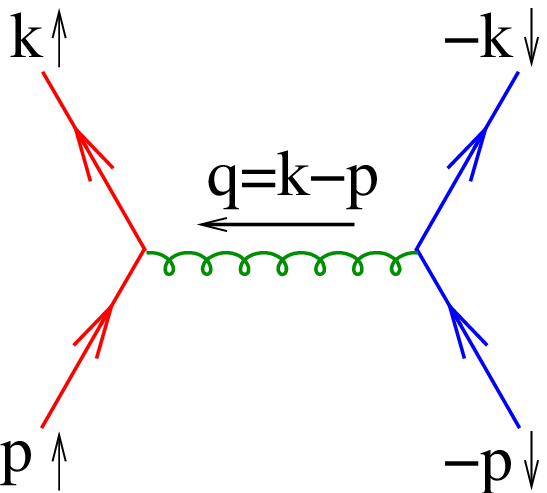}}
\hspace{1em} &
\includegraphics[width=.5\linewidth]{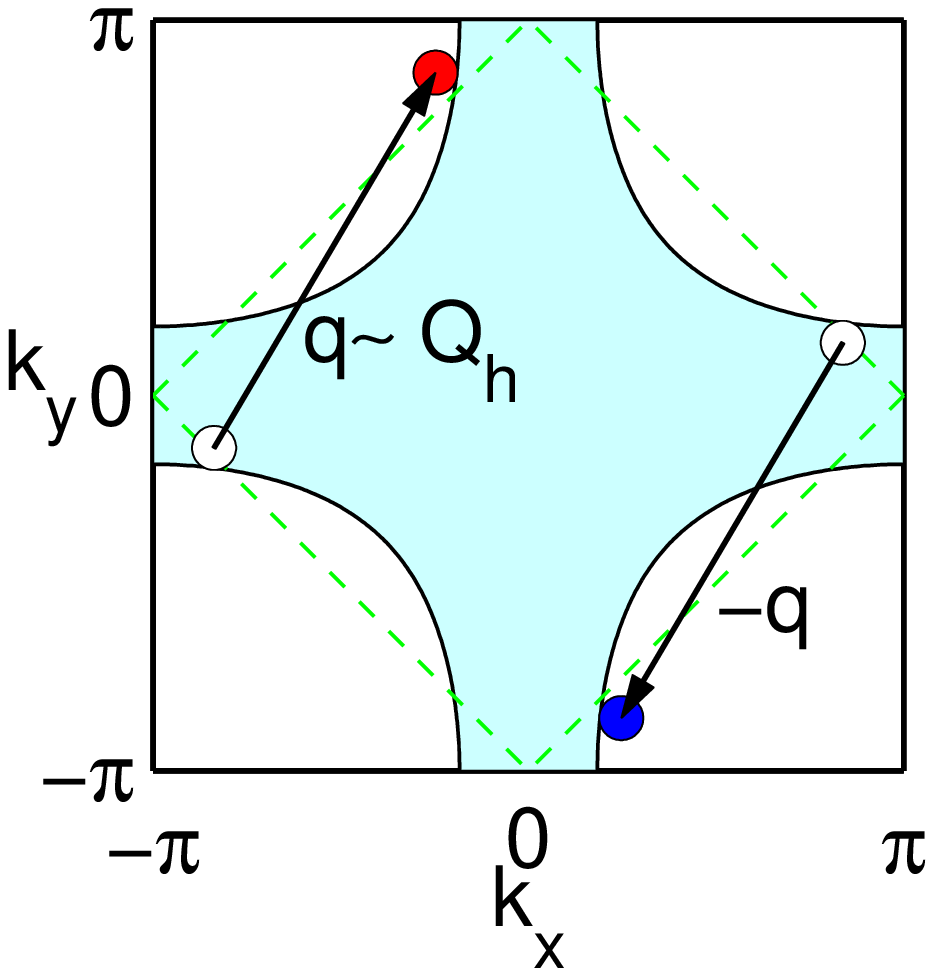}
\end{tabular}
\caption{Effective attraction between spin-broken electrons induced by
    longitudinal spin-orbit fluctuation $\varrho$. The arrows indicate
    spin. The free electron fermi surface is
    plotted for simplicity. The off-diagonal
    regions for dSC or $d_{x^2-y^2}$-SO need not to change
    qualitatively from the state of $d_{xy}$-SO.}
\label{fig:sc_pairing}
\end{figure}

In the effective theory of $d_{xy}$-SO, fermions are coupled to the
longitudinal spin-orbit fluctuation $\varrho$,
\be\textstyle
H_\mathrm{int} = -2 V_B \sum_{\Bk\Bq} [\Gamma^B_{\Bk}+\Gamma^B_{\Bk+\Bq}] 
(\psi^\dag_\Bk \sig_3 \psi_{\Bk-\Bq}) \varrho_\Bq \,. \label{eq:ffrho}
\ee
Broken spin symmetry implies that the $\varrho$ propagator has the form
$(\omega^2 - \SE_\varrho(\Bq)^{2} +i\gamma(\Bq, \omega)^2)^{-1}$ 
with $\SE_\varrho(\Bq)$ having 
minima at $\Bq \rightarrow \pm \BQ_{h,v}$ and the damping rate $\gamma(\Bq,
\omega) \rightarrow 0 $ as $\omega\rightarrow 0$. 
Given the
coupling (\ref{eq:ffrho}), 
we have calculated the effective interaction between fermions
induced by the longitudinal SO fluctuation (Fig.~\ref{fig:sc_pairing}).
In the low energy (static)  limit, the spin-singlet component of the 
Cooper channel interaction is described by in the
Hamiltonian
\be \textstyle
H^{S=0}_\mathrm{int}  = \sum_{\Bk\Bp} v_{\Bk\Bp}
 [\eps_{\rho\sig} \psi^\dag_{\Bk\rho}  \psi^\dag_{-\Bk \sig }]
[ \Bk\rightarrow \Bp]^\dag
\label{eq:HS=0pair}
\ee
where
$
v_{\Bk\Bp}={2g_\varrho V^2_B}  
{[\Gamma^B_\Bk +\Gamma^B_{\Bp}]^2 \SE^{-2}_\varrho(\Bk-\Bp)} >0 
$ and the spin indices $\rho,\sig=\up,\down$.
Here $g_\varrho$ is
a positive normalization constant of the SO field.
The interaction is  repulsive in both
$s$-wave and $d_{xy}$-wave pairing. 
Note that $v_{\Bk\Bp}$, as function of $\Bk-\Bp$, 
peaks at the incommensurate wavevectors $\pm\BQ_{h,v}$
(Fig.~\ref{fig:sc_pairing}). By examining a typical superconducting gap
equation, $\Del^\mathrm{sc}_\Bk =- \sum_\Bp v_{\Bk\Bp}
\Del^\mathrm{sc}_\Bp[(\eps_\Bp-\mu)^2
  +{\Del^\mathrm{sc}_\Bp}^2]^{-1/2}$, 
one finds
nontrivial solution if $\Del^\mathrm{sc}_\Bk$ and
$\Del^\mathrm{sc}_\Bp$ 
have opposite
sign. 
By the  arguments similar to Ref.~\cite{Pines:92}, we conclude that
the favored
superconducting pairing is $d_{x^2-y^2}$-wave. 
Fermions near $(\pm\pi, 0)$ and $(0,\pm \pi)$
will pair as:
$
\avg{\psi_{\Bk\rho} \psi_{-\Bk\sig}} \sim \eps_{\rho\sig}\Gamma^A_\Bk \neq 0\,.
$
The
$\psi$-fermion pair condensate is equivalent to an electron
pair condensate
\be
\eps_{\rho\sig}\avg{\psi_{\Br\rho} \psi_{\Br'\sig}} =
\avg{c^T_\Br U^*_{\Br} i\sig_2 U^\dag_{\Br'} c_{\Br'}}
=\eps_{\rho\sig}\avg{c_{\Br\rho} c_{\Br'\sig}}
\ee
for uniform $\zeta_\Br$.
Thus an exotic normal state can be associated with conventional 
electron pairing.

The state of $d_{x^2-y^2}$-SO is different. A similar coupling
involving the longitudinal SO fluctuation in the $d_{x^2-y^2}$
leads to an effective interaction
that disfavors the $d_{x^2-y^2}$-SC pairing.

$d_{xy}$-SO order does not forbid the existence of Fermi nodal points
in the dSC state. It splits the
quasiparticle energy band around the four triangles along the zone
diagonals (Fig.~\ref{fig:mom_phase_sep}), with Fermi surfaces being
carved out from the lowest possible band.


\paragraph{Incommensurate spin excitations}

The translational symmetry broken by the SO orderings leads to
multi-bands of orbital magnons in a reduced zone scheme, each having
the degeneracy of two transverse spin modes.  The orbital magnons
associated with the $d_{xy}$-SO are gapless at the four incommensurate
momenta $\pm \BQ_h$ and $\pm \BQ_v$.  With coexisting $d_{xy}$ and
$d_{x^2-y^2}$ SO orders, there will be another band that is gapless at
$\Bpi$.  Dynamical spin-spin correlation functions are affected by
particle-hole pair mixing with two orbital magnon channels.  Combining
magnons from $d_{xy}$ and $d_{x^2-y^2}$ can yield two branches of spin
excitation resonance at finite energies, one dispersing upward and
another downward from momentum $\Bpi$. The two branches cross at
$\Bpi$, with zero gap energy.  This structure seems qualitatively
consistent with the inelastic neutron scattering experiments in YBCO
(e.g, as in D. Reznik, et al.~\cite{Reznik:03pre} and references
therein; see also \cite{Mook:01}).  A full investigation is
in progress.


\paragraph{Anisotropic spin susceptibility}
The state of SO ordering has $\avg{\vec{S}_\Br}=0$, 
so the static correlation of {\em local} spins does not exhibit 
conventional long range
order.  A straightforward exercise confirms $\avg{\vec{S}_{\Br}\cdot
\vec{S}_{\Br'}} = -{3\over 8} |\avg{c^\dag_\Br c_{\Br'}}|^2 +{1\over
8} |\avg{c^\dag_\Br\vec{\sig}c_{\Br'}}|^2 \rightarrow 0$ as
$|\Br-\Br'|\rightarrow \infty$.

\begin{figure}[htbp]
\begin{center}
\includegraphics[width=\linewidth]{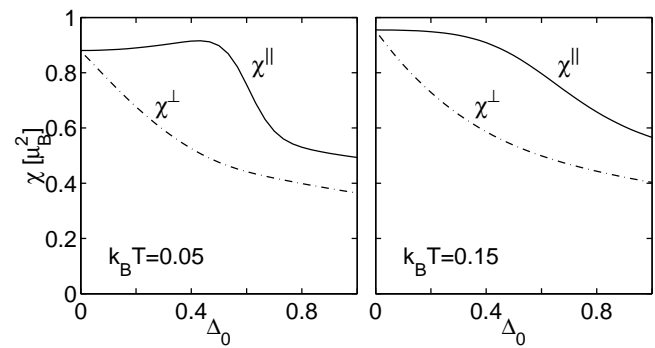}
\end{center}
\caption{Anisotropic spin susceptibility in the $d_{x^2-y^2}$-SO state
     at different temperatures.  Energy units are chosen such that
     $t=1$. Other parameters: $t'=-0.45$ and $\mu=-1.2$ (chemical
     potential). Integration was performed numerically using a mesh of
     $500\times500$ in each quarter of the Brillouin
     zone. $\chi^{\parallel,\perp}$ denote the susceptibilities
     parallel or perpendicular to the direction of the $d_{x^2-y^2}$
     SO. }
\label{fig:spin_chi}
\end{figure}

We have calculated the uniform, static spin susceptibility for the
$d_{x^2-y^2}$-SO 
state, using the effective Lagrangian (\ref{eq:Lpsi}).
We find that the susceptibility is anisotropic,
\be
\begin{array}{rcl}
\chi^{ab} &=& - \mu_B^2 \int {d^2\Bk \over (2\pi)^2} [f'(E^+_\Bk) +
    f'(E^-_\Bk)] \\
&& \hspace{0.1em}
    \times \big[1 + {2(\Del^a_\Bk\Del^b_\Bk - \del^{ab} |\vec{\Del}_\Bk|^2) /
    (E^+_\Bk-E^-_\Bk) }\big]
\end{array}
\ee
where
$ f(E) = {1\over e^{\beta E}+1}$, $f'(E)=\partial f/\partial E$,
   and  the ordering direction is
assumed arbitrary (cf. Eq.~(\ref{eq:Lpsi})).
Fig.~\ref{fig:spin_chi} shows the anisotropy of susceptibility
predicted by the mean
field theory (without corrections due to orbital magnon scatterings).
Measurement of
$\chi^\parallel$ and $\chi^\perp$ could therefore provide important tests of
our proposals. The result is quite different from the AF state, for
which $\chi^{\parallel}< \chi^{\perp}$.


The authors are grateful for comments and advices by 
C. Honerkamp,
P. A. Lee, T. Senthil, S. Weinberg and X.-G. Wen.  This work is
supported in part by funds provided by the U.S. Department of Energy
(D.O.E.) under cooperative research agreement \#DF-FC02-94ER40818.


\bibliography{spin_orbital}

\begin{thebibliography}{13}
\expandafter\ifx\csname natexlab\endcsname\relax\def\natexlab#1{#1}\fi
\expandafter\ifx\csname bibnamefont\endcsname\relax
  \def\bibnamefont#1{#1}\fi
\expandafter\ifx\csname bibfnamefont\endcsname\relax
  \def\bibfnamefont#1{#1}\fi
\expandafter\ifx\csname citenamefont\endcsname\relax
  \def\citenamefont#1{#1}\fi
\expandafter\ifx\csname url\endcsname\relax
  \def\url#1{\texttt{#1}}\fi
\expandafter\ifx\csname urlprefix\endcsname\relax\def\urlprefix{URL }\fi
\providecommand{\bibinfo}[2]{#2}
\providecommand{\eprint}[2][]{\url{#2}}

\bibitem[{ARP()}]{ARPES:rev:03}
\bibinfo{note}{For recent reviews, see, A. Damascelli, Z. X. Shen, and Z.
  Hussain, in Rev. Mod. Phys. 75, 473 (2003); J. C. Campuzano, M. R. Norman,
  and M. Randeria, cond-mat/0209476}.

\bibitem[{\citenamefont{Schulz}(1990)}]{Schulz:90}
\bibinfo{author}{\bibfnamefont{H.~J.} \bibnamefont{Schulz}},
  \bibinfo{journal}{Phys. Rev. Lett.} \textbf{\bibinfo{volume}{64}},
  \bibinfo{pages}{1445} (\bibinfo{year}{1990}).

\bibitem[{\citenamefont{Littlewood et~al.}(1993)\citenamefont{Littlewood,
  Zaanen, Aeppli, and Monien}}]{Littlewood:93}
\bibinfo{author}{\bibfnamefont{P.~B.} \bibnamefont{Littlewood}},
  \bibinfo{author}{\bibfnamefont{J.}~\bibnamefont{Zaanen}},
  \bibinfo{author}{\bibfnamefont{G.}~\bibnamefont{Aeppli}}, \bibnamefont{and}
  \bibinfo{author}{\bibfnamefont{H.}~\bibnamefont{Monien}},
  \bibinfo{journal}{Phys. Rev. B} \textbf{\bibinfo{volume}{48}},
  \bibinfo{pages}{487} (\bibinfo{year}{1993}), \bibinfo{note}{and references
  therein.}

\bibitem[{rem()}]{remark:dSDW}
\bibinfo{note}{Under different names, the $d_{x^2-y^2}$-SO (but not the
  $d_{xy}$) appeared before as one of possible candidate orders in F. Bouis et
  al., cond-mat/9906369; C. Nayak, Phys. Rev. {\bf B62}, 4880 (2000); A. P.
  Kampf and A. A. Katanin, Phys. Rev. {\bf B67}, 125104 (2003); and C. Wu, W.
  V. Liu, and E. Fradkin, Phys. Rev. {\bf B68}, 115104 (2003). To the best of
  our knowledge, the state of this order, however, was never fully studied. In
  a recent paper [K. Maki, B. D\'{o}ra and A. Virosztek, cond-mat/0306567],
  USDW (unconventional spin-density-wave) was proposed to interpreted
  experiments of pseudogap for high $T_c$ cuprates. Their USDW order features
  {\it local} spin component $S^\pm$ in the $a$-$b$ plane.}

\bibitem[{Aff()}]{Affleck-Marston:88;Hsu+:91}
\bibinfo{note}{I. Affleck and J. B. Marston, Phys. Rev. B {\bf 37}, 3774
  (1988); T. Hsu, J. B. Marston and I. Affleck, {\it ibid.} {\bf 43}, 2866
  (1991).}

\bibitem[{\citenamefont{Chakravarty et~al.}(2001)\citenamefont{Chakravarty,
  Laughlin, Morr, and Nayak}}]{Chakravarty+:01}
\bibinfo{author}{\bibfnamefont{S.}~\bibnamefont{Chakravarty}},
  \bibinfo{author}{\bibfnamefont{R.~B.} \bibnamefont{Laughlin}},
  \bibinfo{author}{\bibfnamefont{D.~K.} \bibnamefont{Morr}}, \bibnamefont{and}
  \bibinfo{author}{\bibfnamefont{C.}~\bibnamefont{Nayak}},
  \bibinfo{journal}{Phys. Rev. B} \textbf{\bibinfo{volume}{63}},
  \bibinfo{pages}{094503} (\bibinfo{year}{2001}).

\bibitem[{\citenamefont{Wen and Lee}(1996)}]{Wen-Lee:96}
\bibinfo{author}{\bibfnamefont{X.-G.} \bibnamefont{Wen}} \bibnamefont{and}
  \bibinfo{author}{\bibfnamefont{P.~A.} \bibnamefont{Lee}},
  \bibinfo{journal}{Phys. Rev. Lett.} \textbf{\bibinfo{volume}{76}},
  \bibinfo{pages}{503} (\bibinfo{year}{1996}).

\bibitem[{\citenamefont{Emery et~al.}(1999)\citenamefont{Emery, Kivelson, and
  Tranquada}}]{Emery-Kivelson-Tranquada:99}
\bibinfo{author}{\bibfnamefont{V.~J.} \bibnamefont{Emery}},
  \bibinfo{author}{\bibfnamefont{S.~A.} \bibnamefont{Kivelson}},
  \bibnamefont{and} \bibinfo{author}{\bibfnamefont{J.~M.}
  \bibnamefont{Tranquada}}, \bibinfo{journal}{Proc. Natl. Acad. Sci.}
  \textbf{\bibinfo{volume}{96}}, \bibinfo{pages}{8814} (\bibinfo{year}{1999}).

\bibitem[{\citenamefont{Zaanen and Nussinov}()}]{Zaanen:00pre}
\bibinfo{author}{\bibfnamefont{J.}~\bibnamefont{Zaanen}} \bibnamefont{and}
  \bibinfo{author}{\bibfnamefont{Z.}~\bibnamefont{Nussinov}},
  \bibinfo{note}{cond-mat/0006193}.

\bibitem[{\citenamefont{Weinberg}(1996)}]{Weinberg:bk96:ch19}
\bibinfo{author}{\bibfnamefont{S.}~\bibnamefont{Weinberg}},
  \emph{\bibinfo{title}{The Quantum Theory of Fields II: Modern Applications}}
  (\bibinfo{publisher}{Cambridge University Press},
  \bibinfo{address}{Cambridge, England}, \bibinfo{year}{1996}),
  chap.~\bibinfo{chapter}{19}.

\bibitem[{\citenamefont{Monthoux et~al.}(1992)\citenamefont{Monthoux, Balatsky,
  and Pines}}]{Pines:92}
\bibinfo{author}{\bibfnamefont{P.}~\bibnamefont{Monthoux}},
  \bibinfo{author}{\bibfnamefont{A.~V.} \bibnamefont{Balatsky}},
  \bibnamefont{and} \bibinfo{author}{\bibfnamefont{D.}~\bibnamefont{Pines}},
  \bibinfo{journal}{Phys. Rev. B} \textbf{\bibinfo{volume}{46}},
  \bibinfo{pages}{14803} (\bibinfo{year}{1992}).

\bibitem[{\citenamefont{{D. Reznik, P. Bourges et al.}}()}]{Reznik:03pre}
\bibinfo{author}{\bibnamefont{{D. Reznik, P. Bourges et al.}}},
  \bibinfo{note}{cond-mat/0307591}.

\bibitem[{\citenamefont{Dai et~al.}(2001)\citenamefont{Dai, Mook, Hunt, and
  Dogan}}]{Mook:01}
\bibinfo{author}{\bibfnamefont{P.}~\bibnamefont{Dai}},
  \bibinfo{author}{\bibfnamefont{H.~A.} \bibnamefont{Mook}},
  \bibinfo{author}{\bibfnamefont{R.~D.} \bibnamefont{Hunt}}, \bibnamefont{and}
  \bibinfo{author}{\bibfnamefont{F.}~\bibnamefont{Dogan}},
  \bibinfo{journal}{Phys. Rev. B} \textbf{\bibinfo{volume}{63}},
  \bibinfo{pages}{054525} (\bibinfo{year}{2001}).

\end{thebibliography}


\end{document}